# Valley-related multiple Hall effect in single-layer VSi$_2$P$_4$

Xiangyu Feng, Xilong Xu, Zhonglin He, Rui Peng, Ying Dai[*], Baibiao Huang, Yandong Ma[*]

School of Physics, State Key Laboratory of Crystal Materials, Shandong University, Shandanan Street 27, Jinan 250100, China

*Corresponding author: daiy60@sdu.edu.cn (Y.D.); yandong.ma@sdu.edu.cn (Y.M.)

**Abstract**

2D materials with valley-related multiple Hall effect are both fundamentally intriguing and practically appealing to explore novel phenomena and applications, but have been largely overlooked up to date. Here, using first-principles calculations, we present that valley related multiple Hall effect can exist in single-layer VSi$_2$P$_4$. We identify single-layer VSi$_2$P$_4$ as a ferromagnetic semiconductor with out-of-plane magnetization and valley physics. Arising from the joint effect of inversion symmetry breaking and time reversal symmetry breaking, the exotic spontaneous valley polarization occurs in single-layer VSi$_2$P$_4$, thus facilitating the observation of anomalous valley Hall effect. Moreover, under external strain, band inversion can occur at only one of the valleys of single-layer VSi$_2$P$_4$, enabling the long-sought valley-polarized quantum anomalous Hall effect, and meanwhile the anomalous valley Hall effect is well preserved.. Our work not only enriches the research on valley-related multiple Hall effect, but also opens a new avenue for exploring valley-polarized quantum anomalous Hall effect.

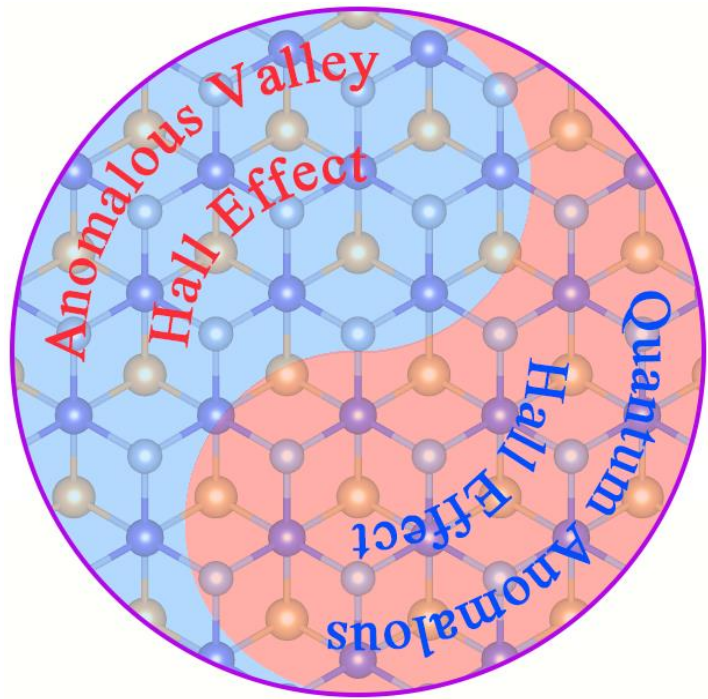

## Introduction

Valley, characterizing energy extrema of conduction or valence band in momentum space, is an emerging degree of freedom of carries in addition to charge and spin [1]. Due to the large separation in momentum space, the valley degree of freedom is robust in terms of low-energy phonons and smooth deformations, which is particularly appealing for information processing. For utilizing valley index as an information carrier, it is crucial to manipulate the carriers in valleys, thereby generating valley polarization [2-5]. Based on the valley-contrasting optical selection rules, optical pumping with circularly polarized light has been employed to produce valley polarization [6-9]. However, optical pumping is a dynamic process, and thus the realized valley polarization is subjected to the lifetime of carriers, making it not applicable for information storage [6-9]. Alternatively, as valleys are related to each other by time-reversal symmetry, the valley degeneracy can also be lifted by breaking time-reversal symmetry, such as magnetic doping[10,11], magnetic proximity effect[12-15], and magnetic field[16-18]. Nevertheless, these external approaches have various problems. For example, magnetic doping tends to form clusters and thus increases impurity scatterings, magnetic proximity effect usually submerges valley physics of host materials, and magnetic field suffers from low efficiency (0.1-0.2 meV/T) [16-18].

The recent rise of 2D ferromagnetic semiconductors provides an unprecedent opportunity for addressing these challenges [19,20]. In 2D ferromagnetic semiconductors with valley physics, the valley polarization occurs spontaneously because of the intrinsic time-reversal breaking, which could facilitate the observation of anomalous valley Hall effect and thus is highly desirable for the field of 2D valleytronics from both fundamental and applied physics. Until now, however, only a few 2D materials have been reported to harbor spontaneous valley polarization, such as 2H-VSe$_2$, LaBr$_2$, Nb$_3$I$_8$, TiVI$_6$, VSi$_2$N$_4$, NbX$_2$ and GdI$_2$[21-28]. Even for these few existing systems, most of them suffer from the in-plane magnetic easy axis in nature [22,24,26,28]; while out-of-plane magnetization is essential for realizing spontaneous valley polarization [23,25,27], the annoying modulation of magnetic easy axis is additionally needed for these systems.

Furthermore, in valley-polarized 2D materials, except for anomalous valley Hall effect, the interplay between valley and other physics is able to enable new phenomena and applications. One exotic example is valley-polarized quantum anomalous Hall effect, which arises from the coupling between topology and valley polarization [29]. Such coupling usually deforms the valley index, giving rise to the absence of anomalous valley Hall effect [30-32]. One the other hand, supposing the valley physics is preserved, the valley-related multiple Hall effect, i.e., anomalous valley Hall effect and quantum anomalous Hall effect, can be realized in one single material simultaneously [33]. This is of significant fundamental and practical importance in condensed matter physics and materials science [34]. Though highly valuable, such valley-related multiple Hall effect in one single material has been rarely explored so far [33-36].

Very recently, Ren et al. successfully synthesized single-layer (SL) $MSi_2N_4$ (M=Mo, W) in experiment via chemical vapor deposition (CVD) [37] and they point out that such CVD method is also applicable for synthesizing many similar 2D materials with a general formula of $MA_2Z_4$, where M represents an early transition metal (W, V, Nb, Ta, Ti, Zr, Hf, or Cr), A is Si or Ge, and Z stands for N, P, or As [37]. Here, we theoretically propose the intriguing valley-related multiple Hall effect can be realized in SL $VSi_2P_4$. SL $VSi_2P_4$ is identified as a ferromagnetic semiconductor with out-of-plane magnetization in nature. Its band edges locate at the +K and -K points, forming a pair of valleys in both the conduction and valence bands. Importantly, due to the breaking of both inversion symmetry and time reversal symmetry, the valley polarization occurs spontaneously in SL $VSi_2P_4$, beneficial for observing the anomalous valley Hall effect. Moreover, by applying strain on SL $VSi_2P_4$, band inversion can be realized at the +K valley, while preserving the band orders trivial at the -K valley, achieving the valley-polarized quantum anomalous Hall effect. Most remarkably, the valley index as well as the anomalous valley Hall effect is preserved under strain, enabling the simultaneously existence of valley-related multiple Hall effect in SL $VSi_2P_4$.

**Computational Methods**

First-principles calculations are performed based on functional theory (DFT) [38-40] as implemented in Vienna ab initio Simulation Package (VASP). The generalized gradient approximation (GGA) in the form of Perdew-Burke-Ernzerhof (PBE) functional is used to treat the exchange-correlation interactions [41]. To describe the strong correction effect for the localized 3d electrons of V atoms, an effective on-site Hubbard term [26,42] of U = 3 eV is utilized. The cutoff energy is set to 500 eV. The vacuum space is set to 18 Å to avoid adjacent interactions. All structures are fully relaxed until the atomic force on each atom is less than 0.01 eV/ Å, and the convergence criterion of the total energy is set to $10^{-5}$ eV. The 11 × 11 × 1 Monkhorst-Pack k-point mesh is used to sample the Brillouin zone. Berry curvature and edge states are calculated by employing the WANNIER90 package [43].

**Results and Discussion**

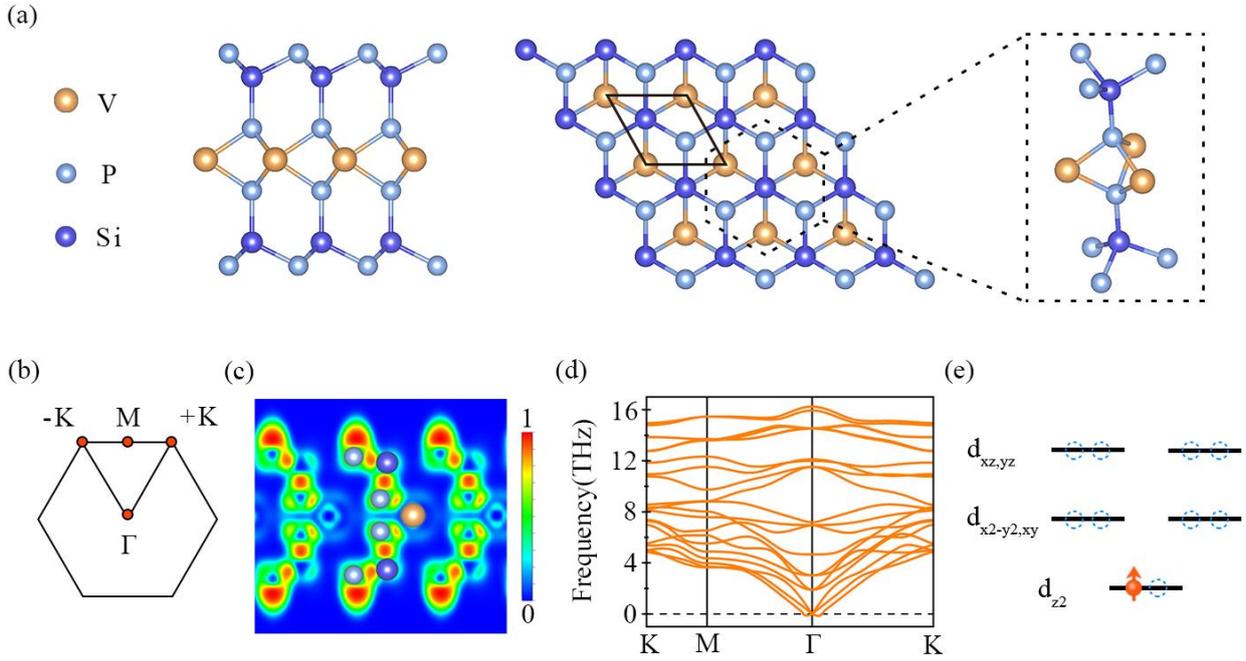

**Fig. 1.** (a) Crystal structure of SL VSi$_2$P$_4$ from top and side views. The solid line indicates the primitive cell and inset in (a) shows the local structure containing two vertical Si atoms and their ligands. (b) 2D Brillouin zone with marking the high-symmetry points. (c) Electron localization function of SL VSi$_2$P$_4$. (d) Phonon spectra of SL VSi$_2$P$_4$. (e) The splitting of d orbitals under the trigonal prismatic crystal field.

The crystal structure of SL VSi$_2$P$_4$ is shown in **Fig. 1(a)**. It exhibits a hexagonal lattice with the space group of P$\bar{6}$m2. Each unit cell contains one V, two Si and four P atoms, which are stacked in the sequence of P-Si-P-V-P-Si-P. Each V atom is coordinated with six P atoms, forming a trigonal prismatic configuration. Similar with the H phase of MoS$_2$, the inversion symmetry of SL VSi$_2$P$_4$ is broken. The lattice constant of SL VSi$_2$P$_4$ is optimized to 3.476 Å, agreeing well with previous work [44]. To investigate the bonding characteristics of SL VSi$_2$P$_4$, the electron localization function (ELF) is calculated. As shown in **Fig. 1(c)**, for the Si-P bond, electrons are mainly localized at center area between them, indicating a covalent bonding. While for the V-P bond, electrons are highly localized around P and V atoms, suggesting an ionic bonding. To confirm the stability of SL VSi$_2$P$_4$, the phonon spectra is calculated. As shown in **Fig. 1(d)**, except for the tiny imaginary frequencies around the Γ point, all branches are positive, which confirms that SL VSi$_2$P$_4$ is dynamically stable.

Under the trigonal prismatic crystal field, the d orbitals of V atom split into three group: one *a* orbital (d$_{z2}$), two *e$_1$* orbitals (d$_{xy}$, d$_{x2-y2}$) and two *e$_2$* orbitals (d$_{xz}$, d$_{yz}$), see **Fig. 1(e)**. The valence electron configuration of V atom is $3d^34s^2$. After denoting four electrons to the neighboring P atoms, one valence electron is left. According to the Hund's rule and Pauli exclusion principle, the electron configuration of $V^{4+}$ is $a^1e_1^0e_2^0$, see **Fig. 1(e)**. Therefore, a magnetic moment of 1 $\mu_B$ per unit cell is expected for SL $VSi_2P_4$. Our spin-polarized calculations indeed show that SL $VSi_2P_4$ is spin-polarized, and the magnetic moment per unit cell is 1 $\mu_B$. The magnetic moment is mainly distributed on the V atom, with a slight distribution on the neighboring P atoms. To determine the magnetic ground state of SL $VSi_2P_4$, three magnetic configurations are considered, including one ferromagnetic (FM) state, one antiferromagnetic (AFM) state and one nonmagnetic (NM) state. The FM configuration is found to be the magnetic ground state for SL $VSi_2P_4$, which is 75 and 477 meV lower in energy than the AFM and NM configurations, respectively. The FM ground state in SL $VSi_2P_4$ can be understood by examining the crystal structure. In SL $VSi_2P_4$, the V-P-V bonding angle is estimated to be 91.7°, close to 90.0°. According to Goodenough-Kanamori-Anderson rules[45-47], such a structure configuration favors FM coupling.

Based on the Mermin-Wagner theorem [48], stable FM coupling is forbidden in 2D isotropic Heisenberg model. However, finite magnetic anisotropy is able to protect long-range magnetic order. Therefore, the stability of FM order in SL $VSi_2P_4$ strongly correlates with its magnetic anisotropy energy (MAE), which is defined as the energy difference between in-plane and out-of-plane magnetization: MAE = $E_{x/y}$-$E_z$. The MAE of SL $VSi_2P_4$ is calculated to be 6.38 μeV per unit cell, suggesting that its easy axis is along the out-of-plane direction. This finite value is larger than those of Ni, Fe and Co with MAE on the order of 1 μeV per unit cell [49]. Moreover, such intrinsic out-of-plane magnetization is tantalizing for realizing the spontaneous valley polarization that we will show later. In previous reported 2D ferromagnetic materials with spontaneous valley polarization, most of them suffer from in-plane magnetization in nature, and additional tuning of magnetization from in-plane to out-of-plane is necessarily needed [22,24,26,28]]. Apparently, such an additional tuning using external approaches is rather challenge [15,50,51].

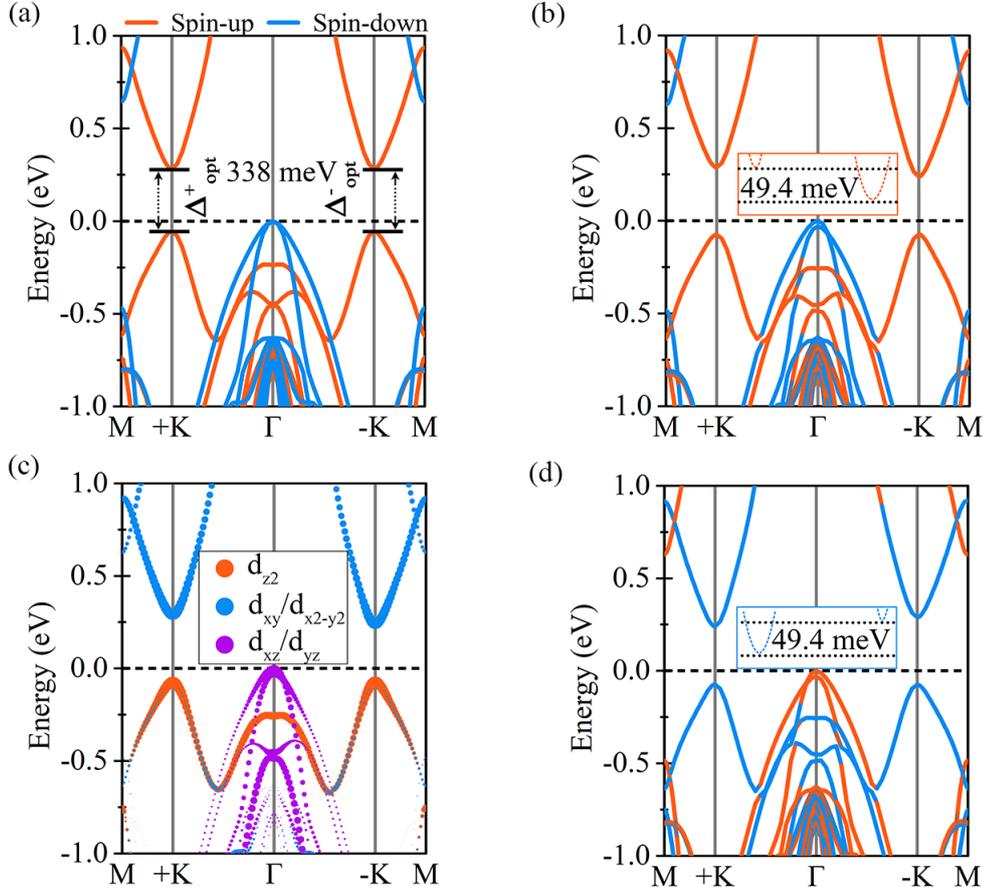

**Fig. 2**. Spin-polarized band structure of SL VSi$_2$P$_4$ (a) without and (b) with considering SOC. (C) Orbital-resolved spin-polarized band structure of SL VSi$_2$P$_4$ with considering SOC. (d) is same as (b), but with opposite magnetization. The Fermi level is set to 0 eV.

**Fig. 2(a)** presents the spin-polarized band structure of SL VSi$_2$P$_4$ without considering SOC. Because of the intrinsic spin polarization, the spin-up and spin-down channels split significantly. SL VSi$_2$P$_4$ exhibit a semiconducting character with an indirect band gap of 0.28 eV. From **Fig. 2(a)**, it can be seen that its conduction and valence band edges are from opposite spin channels. Therefore, SL VSi$_2$P$_4$ is a bipolar ferromagnetic semiconductor. Furthermore, as shown in **Fig. 2(a)**, its conduction band minimum (CBM) locates at the +K and -K points, forming a pair of degenerate valleys in the conduction band. While for its valence band, although the valence band maximum (VBM) lies at the Γ point, local extrema at the +K and -K points are also formed, endowing SL VSi$_2$P$_4$ with a pair of degenerate valleys in the valence band as well. Both of these two pairs of valleys are from the spin-up channels. Moreover, because of broken inversion symmetry, the degenerate valleys at the +K and -K points in both the conduction and valence bands are inequivalent. As a result, SL VSi$_2$P$_4$ is a 2D valleytronic semiconductor.

When taking SOC into consideration, as shown in **Fig. 2(b)**, the +K valley shifts above the -K valley in energy in the conduction band. While for the valence band, the +K valley down-shifts below the -K valley. This indicates that the valley degeneracy in both the conduction and valence bands are lifted, giving rise to the intriguing valley polarization. It should be noted that the valley polarization in SL VSi$_2$P$_4$ occurs spontaneously, without needing any additional tuning. This is particularly interesting for practical valleytronic applications. The spontaneous valley polarization in the conduction band is found to be 49.4 meV, while it is only 3 meV in the valance band. When reversing the magnetization orientation, as shown in **Fig. 2(d)**, the valleys are from the spin-down channels, and the valley polarization in both the conduction and valance bands are reversed.

The underlying physics for the spontaneous valley polarization in SL VSi$_2$P$_4$ can be attributed to the combined effect of SOC and magnetic exchange interaction. When excluding SOC but considering magnetic exchange interaction, due to the time-reversal symmetry breaking, the spin splitting for the bands at the +K and -K valleys are $\Delta_{mag}^{+K}= \Delta_{mag}^{-K}$ ($\Delta^{+K/-K}= E_\uparrow^{+K/-K} - E_\downarrow^{+K/-K}$, where ↑ and ↓ represent spin-up and spin-down states, respectively). On the other hand, when including SOC but excluding magnetic exchange interaction, the spin splitting also occurs at the +K and -K valleys. However, because of the broken inversion symmetry and time-reversal symmetry protection, the absolute values for the valley splitting at the +K and -K valleys are the same, but the signs are opposite, namely, $\Delta_{SOC}^{+K} = -\Delta_{SOC}^{-K}$. As a consequence, when considering both SOC and magnetic exchange interaction, the net spin splitting are $\Delta_{mag}^{+K} + \Delta_{SOC}^{+K}$ at the +K valley and $\Delta_{mag}^{-K} + \Delta_{SOC}^{-K}$ at the -K valley, leading to the intriguing spontaneous valley polarization.

As mentioned above, the valley polarization in the conduction band of SL VSi$_2$P$_4$ is as large as 49.4 meV, while that in the valence band almost can be neglected. This discrepancy is related to the fact that valleys in the conduction and valence bands are contributed by different orbitals. From the orbital-resolved band structure of SL VSi$_2$P$_4$ shown in **Fig. 2(c)**, it can be seen that the valleys in conduction band is dominated by V-$d_{x^2-y^2/xy}$ orbitals, while the valleys in valence band is mainly from V-$d_{z^2}$ orbitals. As SOC plays an important role in the valley polarization of SL VSi$_2$P$_4$, to get underlying physics for the discrepancy in valley polarization, we employ a simple model to describe the SOC term as follows [52,53]:

$$\hat{H}_{SOC} = \lambda \hat{L} \cdot \hat{S} = \hat{H}_{SOC}^0 + \hat{H}_{SOC}^1$$

Here, $\hat{L}$ and $\hat{S}$ are orbital angular and spin angular operators, respectively. $\hat{H}^0$ and $\hat{H}^1$ indicate the interactions between same spin states and between opposite spin states, respectively. From the band structures shown in **Fig. 2**, it can be seen that the states around the valleys consist of only one spin channel, while the states from the other spin channel lie far away from the valleys. Therefore, we can ignore $\hat{H}^1$ here. Concerning $\hat{H}^0$, it can be expressed in terms of polar angles of $\theta$ and $\phi$[54,55]:

$$\hat{H}^0 = \lambda \hat{S}_{z'} \left( \hat{L}_z \cos\theta + \frac{1}{2}\hat{L}_+ e^{-i\phi}\sin\theta + \frac{1}{2}\hat{L}_- e^{+i\phi}\sin\theta \right)$$

Considering the intrinsic out-of-plane magnetization of SL VSi$_2$P$_4$, $\theta = 0$. Then, $\hat{H}^0$ can be simplified as:

$$\hat{H}^0 = \lambda \hat{S}_{z'} \hat{L}_z = \alpha \hat{L}_z$$

Based on the C$_3$ symmetry of the +K and -K valleys, the basis functions can be chosen as:

$$|\varphi_c^\tau\rangle = \sqrt{\frac{1}{2}} \left( |d_{x^2-y^2}\rangle + i\tau |d_{xy}\rangle \right)$$

$$|\varphi_v\rangle = |d_{z^2}\rangle$$

Here $\tau = \pm 1$ indicate the valley index at $\pm K$ point, and $c/v$ represents the conduction/valance band. The energy levels of valleys can be expressed as:

$$E_c^\tau = \langle \varphi_c^\tau | \hat{H}^0 | \varphi_c^\tau \rangle$$

$$E_v = \langle \varphi_v | \hat{H}^0 | \varphi_v \rangle$$

Then, the valley polarization in the conduction and valance bands can be written as:

$$E_c^{-K} - E_c^{+K} = i\langle d_{xy}|\hat{H}^0|d_{x^2-y^2}\rangle - i\langle d_{x^2-y^2}|\hat{H}^0|d_{xy}\rangle = -4\alpha$$

$$E_v^{-K} - E_v^{+K} = 0$$

According to these results, the discrepancy in the valley polarization in conduction and valence bands of SL VSi$_2$P$_4$ can be easily understood.

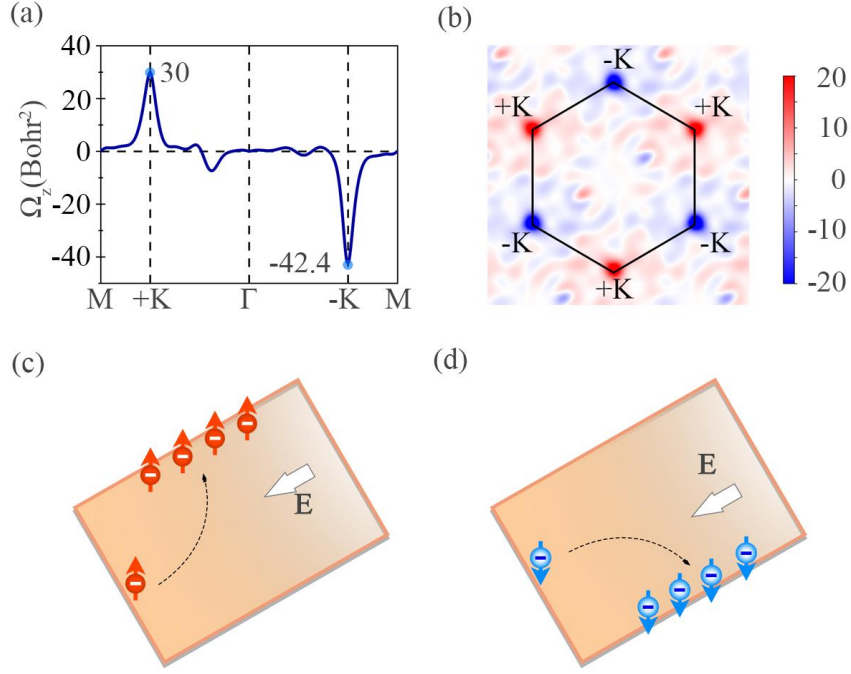

**Fig. 3**. Berry curvature of SL VSi$_2$P$_4$ (a) as a curve along the high-symmetry points and (b) as a counter map over the 2D Brillouin zone. (c, d) Diagrams of the anomalous valley Hall effect under electron doping and an in-plane electric field. (d) is same as (c), but with opposite magnetization.

After estimating the spontaneous valley polarization, we investigate the valley-contrasting physics in SL VSi$_2$P$_4$. To this end, we calculate the Berry curvature $\Omega$ of SL VSi$_2$P$_4$, which is defined as [56]:

$$\Omega(k) = -\sum_n \sum_{n\neq n'} f_n \frac{2Im\langle\psi_{nk}|v_x|\psi_{n'k}\rangle\langle\psi_{nk}|v_y|\psi_{n'k}\rangle}{(E_n - E_{n'})^2}$$

Here, $f_n$, $E_n(k)$ and $v_{x/y}$ are the Fermi-Dirac distribution function, eigenvalue of the blotch state $\psi_{nk}$, and velocity operator, respectively. **Fig. 3(a, b)** plots the calculated Berry curvature of SL VSi$_2$P$_4$. Obviously, two peaks of the Berry curvature, respectively, locate at the +K and -K valleys. The Berry curvature at the +K and -K valleys have opposite signs and different absolute values. The former character arises from the broken inversion symmetry of SL VSi$_2$P$_4$, while the later corelates with the broken time reversal symmetry. Therefore, SL VSi$_2$P$_4$ shows a valley-contrasting Berry curvature. Under an in-plane electric filed $E$, the Berry curvature can behave like an effective magnetic field for the carriers, endowing the carrier with an anomalous velocity, $v \sim E \times \Omega(k)$ [57]. Based on this fact, the anomalous valley Hall effect can be realized in SL VSi$_2$P$_4$. As illustrated in **Fig. 3(c)**, when shifting the Fermi level between the +K and -K valleys in the conduction band via electron doping, the spin-up electrons from the -K valley will acquire an anomalous velocity and accumulate on the upper side of the sample in the presence of an in-plane electric field, realizing the anomalous valley Hall effect. Upon reversing the magnetization orientation of SL VSi$_2$P$_4$, as shown in **Fig. 3(d)**, when sifting the Fermi level between the +K and -K valleys in the conduction band, the spin-down electrons from the +K valley will gather to the lower boundary of the sample under an in-plane electric field. Interestingly, in addition to the anomalous valley Hall effect, the charge and spin Hall effect are also realized in SL VSi$_2$P$_4$, see **Fig. 3(c,d)**. In this case, the valley degree of freedom can be manipulated by magnetization and read out by electric measurement, facilitating the application of valley degree of freedom in information processing.

Epitaxial strain is known as an effective mean for the manipulation of electronic and magnetic properties of 2D materials. In the following, we study the electric and magnetic responses of SL VSi$_2$P$_4$ by applying in-plane biaxial strain. Here, in-plane strain is defined as $\varepsilon = (a - a_0)/a_0$, where $a(a_0)$ is the lattice constant of SL VSi$_2$P$_4$ with (without) strain. As shown in **Fig. 4(a)**, the FM ground state for SL VSi$_2$P$_4$ is stable under stain. **Fig. S1** presents the band structures of SL VSi$_2$P$_4$ under various strains. It can be seen that under tensile strain, the indirect band gap character is preserved. With increasing tensile strain, the energy difference between the valleys and VBM at the Γ point increases. For the valley polarization, the value changes slightly under tensile strain. This can be attributed to the fact that the valley polarization mainly relates to SOC strength, which is intrinsic physics and robust against strain.

With increasing compressive strain, as shown in **Fig. S1**, the valleys in the valence band shift downwards with respect to VBM at the Γ point, and the valleys in the conduction band shift upwards with respect to the lowest conduction band at the M point. This results in that the band gap of SL VSi$_2$P$_4$ first transfers into direct under strain of -1%, then into indirect, and then is closed under strain of -5%. Interestingly, by looking at the band structure of SL VSi$_2$P$_4$ under -3% strain, the large valley polarization occurs in the valence band, while it is negligible for the valley polarization in the conduction band. This is in sharp contrast to the pure case where the valley polarization in the conduction band is significantly larger than that in the valence band. Therefore, by utilizing strain, the magnitude of valley polarization can be switched between the conduction and valence bands, which is highly desirable for controllable valleytrinic applications.

As mentioned above, the magnitude of valley polarization in SL VSi$_2$P$_4$ corelates with orbital contributions. Apparently, as compared with the pure case, the orbital contributions for the valleys in the conduction and valence bands are switched under -3% strain. Namely, band inversion between V-$d_{x^2-y^2}/xy$ and V-$d_{z^2}$ orbitals occur at both the +K and -K valleys. To monitor the band inversion process, we present more band structures of SL VSi$_2$P$_4$ under the strain between -2% and -3%. As shown in **Fig. S2**, when increasing compressive strain to -2.02%, the band gap is closed at the -K valley, and a narrow band gap still exists at the +K valley, giving rise to the exotic half-valley-metal state [34]. When further increasing the compressive strain, band inversion occurs at the -K valley, while the band orders at the +K valley remains trivial. In this case the valley-polarized quantum anomalous Hall state is realized, which we will confirm later. When increasing the compressive strain to -2.30%, the band gap at the -K valley is maintained, but the band gap at the +K valley is closed, leading to the other half-valley-metal state. With further increasing the compressive strain larger than -2.30%, band inversion appears at both the +K and -K valleys, and trivial topology is expected.

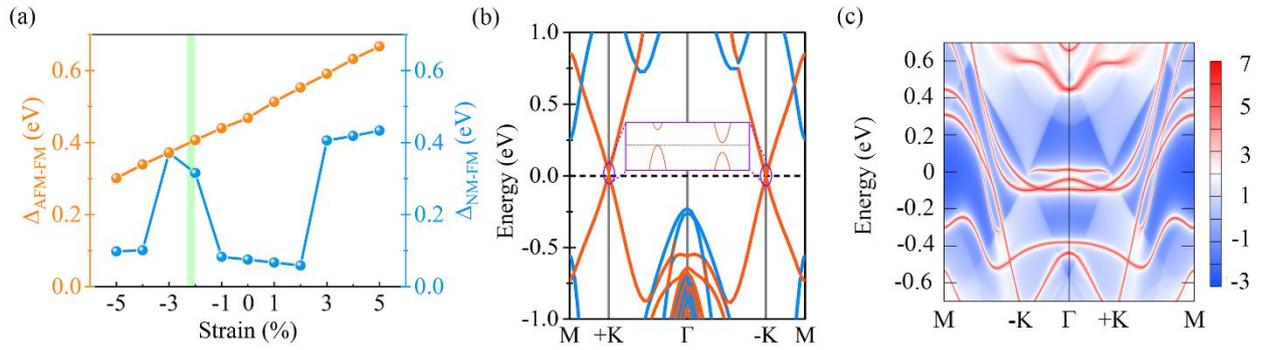

**Fig. 4** (a) Energy difference between different magnetic configurations [$\Delta_{AFM-FM}$: energy difference between AFM and FM configurations; $\Delta_{NM-FM}$: energy difference between NM and FM configurations] as a function of strain. (b) Band structure and (c) edge state of SL VSi$_2$P$_4$ under -2.1% strain. Insert in (b) shows the enlarged band edges around the +K and -K points. The Fermi level is set to 0 eV.

To confirm the strain-induced valley-polarized quantum anomalous Hall effect in SL VSi$_2$P$_4$, we take the case under 2.1% compressive strain as an example to investigate the Chen number. Our calculations show that the Chen number integrated over the Berry curvatures is nonzero (C = 1), which confirms the nontrivial topological state of SL VSi$_2$P$_4$. As topological edge state is the unique fingerprint of quantum anomalous Hall state, we also calculate the edge state of SL VSi$_2$P$_4$ under 2.1% compressive strain. As shown in **Fig. 4(c)**, there is an edge state connecting the valence and conduction bands, consisting with the Chen number calculations. Therefore, in addition to the anomalous valley Hall effect, the valley-polarized quantum anomalous Hall effect as also realized in SL VSi$_2$P$_4$, offering a wonderful platform to explore interplay between valley and topology physics.

**Conclusion**

In summary, using first-principles calculations, we report our identification of valley-related multiple Hall effect in SL VSi$_2$P$_4$. We find that SL VSi$_2$P$_4$ is a ferromagnetic semiconductor with out-of-plane magnetization and valley physics. Arising from the broken inversion symmetry and broken time-reversal symmetry, the valley polarization occurs spontaneous in SL VSi$_2$P$_4$, which facilitates the observation of anomalous valley Hall effect as well as the charge and spin Hall effect. Furthermore, under compressive strain, the band inversion can be realized at the -K valley, while retaining the trivial band orders at the +K valley, giving rise to the valley-polarized quantum anomalous Hall effect. Remarkably, the valley physics is preserved under compressive strain, indicating the valley-related multiple Hall effect can simultaneously exist in SL VSi$_2$P$_4$. All these findings together with the recent progress in SL MoSi2N4 point out an experimentally achievable scheme for exploring valley-related multiple Hall effect in 2D lattice.

**Acknowledgement**

This work is supported by the National Natural Science Foundation of China (Nos. 11804190 and 12074217), Shandong Provincial Natural Science Foundation (Nos. ZR2019QA011 and ZR2019MEM013), Shandong Provincial Key Research and Development Program (Major Scientific and Technological Innovation Project) (No. 2019JZZY010302), Shandong Provincial Key Research and Development Program (No. 2019RKE27004), Shandong Provincial Science Foundation for Excellent Young Scholars (No. ZR2020YQ04), Qilu Young Scholar Program of Shandong University, and Taishan Scholar Program of Shandong Province.

LUMO interactions using spin-orbit coupling as perturbation. *Acc. Chem. Res.* 48, 3080 (2015).

[55] M.-H. Whangbo, H. Xiang, H.-J. Koo, E. E. Gordon, and J. L. Whitten, Electronic and structural factors controlling the spin orientations of magnetic ions. *Inorg. Chem.* 58, 11854 (2019).

[56] D. J. Thouless, M. Kohmoto, M. P. Nightingale, and M. den Nijs, Quantized Hall Conductance in a Two-Dimensional Periodic Potential. *Phys. Rev. Lett.* 49, 405 (1982).

[57] D. Xiao, M.-C. Chang, and Q. Niu, Berry phase effects on electronic properties. *Rev. Mod. Phys.* 82, 1959 (2010).

**Supporting Information**

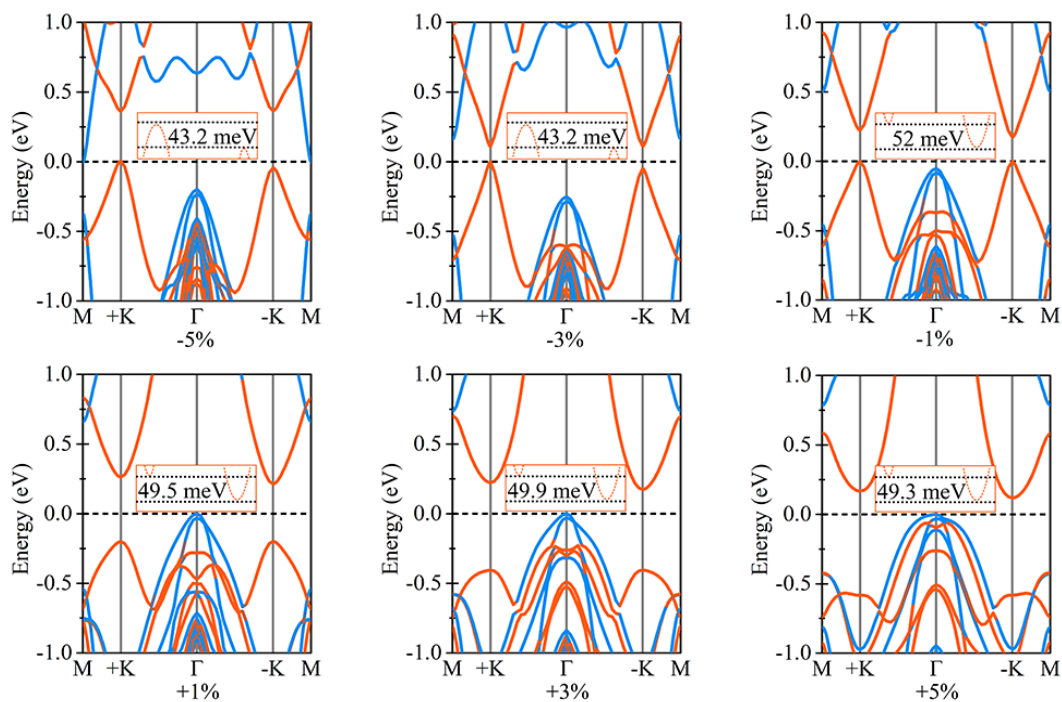

**Fig. S1** Spin-polarized band structures of SL VSi$_2$P$_4$ with considering SOC under various strains. Orange and blue lines represent the spin-up and spin-down states, respectively. The Fermi level is set to 0 eV.

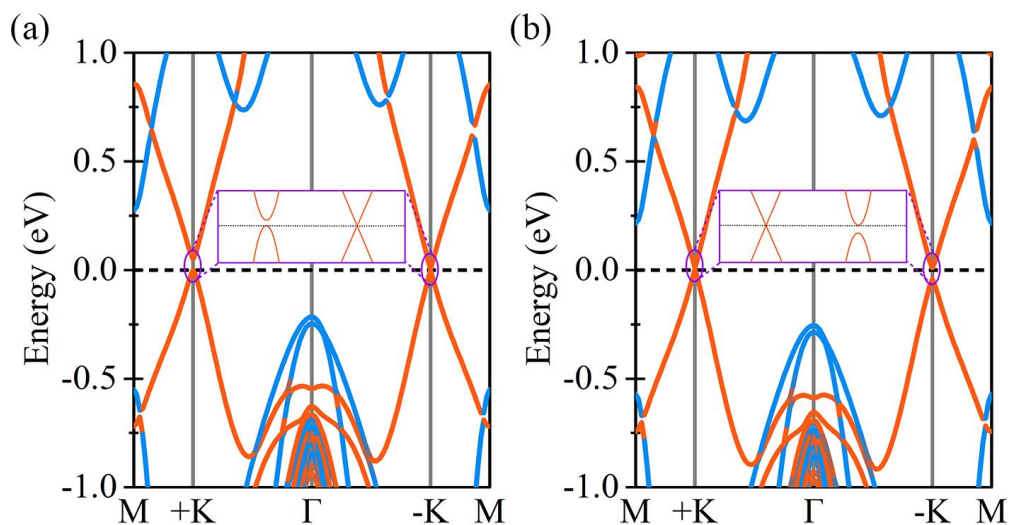

**Fig. S2** Band structures of SL Cr$_2$Se$_3$ under biaxial strains of (a) -2.02% and (b) +2.30%. Inserts show the enlarged band edges around the +K and -K points. The Fermi level is set to 0 eV.